\documentclass[a4paper]{jpconf}
\usepackage{graphicx}
\begin{document}
\title{Building a Hydrodynamics Code with Kinetic Theory}
\author{Irina Sagert$^1$, Wolfgang Bauer$^{1,2}$, Dirk Colbry$^2$, Rodney Pickett$^2$, Terrance Strother$^3$}
\address{$^1$Department of Physics and Astronomy, Michigan State University, East Lansing, Michigan, 48824, USA\\
$^2$Institute for Cyber-Enabled Research, Michigan State University East Lansing, Michigan 48824, USA\\
$^3$XTD-6, Los Alamos National Laboratory, Los Alamos, New Mexico 87545, USA}
\ead{sagert@nscl.msu.edu}

\begin{abstract}
We report on the development of a test-particle based kinetic Monte Carlo code for large systems and its application to simulate matter in the continuum regime. Our code combines advantages of the Direct Simulation Monte Carlo and the Point-of-Closest-Approach methods to solve the collision integral of the Boltzmann equation. With that, we achieve a high spatial accuracy in simulations while maintaining computational feasibility when applying a large number of test-particles. The hybrid setup of our approach allows us to study systems which move in and out of the hydrodynamic regime, with low and high particle densities. To demonstrate our code's ability to reproduce hydrodynamic behavior we perform shock wave simulations and focus here on the Sedov blast wave test. The blast wave problem describes the evolution of a spherical expanding shock front and is an important verification problem for codes which are applied in astrophysical simulation, especially for approaches which aim to study core-collapse supernovae.
\end{abstract}
\section{Introduction}
Hydrodynamics is a frequent approach to describe the evolution of matter via its thermodynamic properties. Hereby, the physical particles the medium is comprised of have mean-free-paths $\lambda$ which are much smaller than a characteristic length scale of the studied system $L$. If the so-called Knudsen number $K_n$ is small with $K_n = \lambda / L \ll 1$, the continuum approximation can be applied and the evolution of matter is described by the Navier-Stokes equations. The latter are a set of coupled differential equations and are derived from the conservation of mass, momentum, and energy density as well as the assumption of local thermal equilibrium \cite{Landau75}. The Navier-Stokes equations can be solved numerically either by the Eulerian, i.e. grid-based, method or the Lagrangian approach, also known as Smooth Particle Hydrodynamics \cite{Gingold77}. Numerical hydrodynamics is a widely applied tool in science and engineering. However, if the mean-free-path of physical particles becomes large, resulting in $K_n \gg 1$, the continuum approximation breaks down. As a consequence, the Navier-Stokes equations cannot be applied anymore to evolve the system and transport equations have to be solved instead.  
\section{Kinetic Theory}
In transport theory the many-body problem is solved via modeling the evolution of the particle phase-space density function $f(\vec{x}, \vec{p}, t)$ with position $\vec{x}$, momentum $\vec{p}$, and time $t$. Hereby, the equation of motion of the n-body density matrix is rewritten into a one-particle evolution, coupled to an equation of motion for the n-body correlation function. Truncation of the latter to the three-body level leaves a system of coupled non-linear equations of motion for the one-particle density function and the two-body correlation function. To derive the transport equations, the Wigner transform is applied to the particle density function and is then expanded to sums over all possible single particle states. Rearranging the equation of motion to a collisionless one-body propagation and collision terms results in the so-called Boltzmann transport equation \cite{Chapman70}:
\begin{eqnarray}
\frac{\partial f}{\partial t} + \frac{\vec{p}}{m} \cdot \vec{\nabla }_r \: f + \vec{F} \frac{\partial f}{\partial \vec{p}} = \left. \frac{\partial f}{\partial t} \right|_{\mathrm{Coll}} .
\label{boltzmann}
\end{eqnarray} 
The left-hand side of eq.(\ref{boltzmann}) represent the collisionless motion of an ensemble of particles in an external force field $\vec{F}$ while the change in the particles' phase space distribution function $f=f(\vec{x}, \vec{p}, t)$ due to binary collisions is given by the collision integral on the right hand side. The probabilities of particle interaction depend hereby on their interaction cross-sections $\sigma$ which are connected to the particle mean-free-paths via $\sigma \propto 1/\lambda$. Numerically, the Boltzmann transport equation can be solved by the test-particle method. Hereby, the phase space distribution function is represented by a sum of delta functions \cite{Wong82}:
\begin{eqnarray}
f(\vec{x}, \vec{p},t) = \sum_{i=0}^{N} \delta^3 \left( \vec{x} - \vec{x}_i (t) \right) \:  \delta^3 \left( \vec{p} - \vec{p}_i (t) \right) ,
\end{eqnarray}    
whereas $N$ is the number of test-particles in the simulation. Inserting $f(\vec{x}, \vec{p},t)$ into eq.(\ref{boltzmann}) results in $2N$ coupled first-order differential equations of motion: 
\begin{eqnarray}
\frac{d}{dt}\vec{p}_i = \vec{F}(\vec{r}_i)+\vec {\cal C}(\vec{p}_i), \: \: \frac{d}{dt}\vec{r}_i = \frac{\vec{p}_i}{m_{i}}, \: \: i = 1,\ldots,N .
\end{eqnarray}
Hereby, $\vec{\mathcal{C}}(\vec{p}_i)$ represents the effects of two-body collisions on the $i^{th}$ test-particle's momentum.\\
Kinetic methods are typically applied for rarefied gases with large Knudsen numbers. Examples for transport models applications are studies of hypersonic flow \cite{Boyd95}, nano-scale devices \cite{Yap12}, particle production in heavy ion collisions \cite{Bauer86,Aichelin86,Bouras10,Kortemeyer95}, the dynamics of inertial confinement fusion (ICF) capsules \cite{Casanova91}, and astrophysics \cite{Hernquist92}. However, application areas can be extended to describe systems that can move in and out of the continuum limit. Especially heavy-ion collisions and core-collapse supernovae contain baryons which are in hydrodynamic equilibrium, as well as particles with mean-free-paths which can be very large. In heavy-ion collisions the latter are e.g. $\pi$ mesons while for core-collapse supernovae, neutrinos can be trapped, free-streaming, or diffusing in the baryonic matter. In core-collapse supernovae, neutrinos are typically evolved by coupling solvers for the neutrino transport equations to the evolution of the hydrodynamics equations. However, while for one-dimensional core-collapse supernova simulations, the Boltzmann transport equation can be solved exactly \cite{Liebendoerfer05}, multi-dimensional calculations are currently computationally not feasible and can only be performed with approximations to the neutrino transport \cite{Liebendoerfer09}. On the other hand, Monte Carlo neutrino transport has been suggested as an alternative approach since it is expected to scale better for multi-dimensional simulations \cite{Adbikamalov12}.\\
Transport codes are able to handle matter in different states, from low density rarefied gases to high density matter in the continuum regime \cite{Alexander95}. However, since the simulation of matter in the continuum regime requires the frequent interaction of particles with each other, large values of $N$ are computationally challenging. One of the most efficient approaches to solve the collision integral is the Direct Simulation Monte Carlo (DSMC) method \cite{Bird70}. Hereby, the simulation space is partitioned into a scattering grid and interaction partners are chosen randomly amongst particles within a cell. The resulting simulation times scale as $O (N \log N)$ and can be reduced even further by performing the calculations in parallel on multiple processors. However, a consequence of the probabilistic choice of interaction partners is that spatial details cannot be resolved to smaller scales than the size of a grid cell. Furthermore, the finite separation between collision partners and the instantaneous character of interactions can lead to causality violations in a relativistic regime \cite{Kortemeyer95}. An approach which minimized the distance between interaction partners is the Point-of-Closest-Approach (PoCA) method. Hereby, scattering partners are determined as test-particles whose path crosses within one timestep. The PoCA method has been successfully applied in modeling physical systems with limited number of constituents, for example the simulation of heavy-ion collisions \cite{Bertsch88,Cugnon81,Stoecker86}. While such simulations have very good spatial resolution, the PoCA method typically requires the comparison of each particle with every other particle of the simulation and has a computational scaling of $O(N^2)$. The time to search for interaction partners can be reduced by a spatial sort of particles. Nevertheless, for large values of $N$, PoCA methods are computationally very expensive. 
\section{Combination of the DSMC and PoCA methods}
In our approach we combine the advantages of the DSMC and the PoCA techniques to solve the collision integral of the transport equations. Our final goal is to study astrophysical systems such as core-collapse supernovae and matter in ICF capsules via the kinetic scheme. Similar to the DSMC method we divide the simulation space into a grid. However, interaction partners are not chosen randomly but are determined from test-particles in a grid cell and the neighboring cells via the PoCA method. With that, we ensure a spatial accuracy which is higher than in usual DSMC simulations. To avoid a computationally expensive spatial sorting we connect and propagate particles in their corresponding cell via a linked lists. Furthermore, the setup of the scattering grid allows us to use multiple processors and determine interaction partners for different cells in parallel. With that, though computationally more expensive than DSMC methods, our algorithm has a much smaller computational time than traditional PoCA methods and can thereby describe systems with a large number of interacting particles. Collision partners are determined via three steps \cite{Sagert2013}. First, possible interaction partners are selected as particles reaching their point of closest approach during the timestep $\Delta t$. For that we determine the sign of the crossing number $\chi$:
\begin{eqnarray}
\chi = ( \vec{r}_{\mathrm{rel}} (t) \cdot \vec{v}_{\mathrm{rel}} (t) ) ( \vec{r}_{\mathrm{rel}} (t + \Delta t) \cdot \vec{v}_{\mathrm{rel}} (t + \Delta t) ) .
\label{c_value}
\end{eqnarray}
Hereby, $\vec{r}_{\mathrm{rel}}$ and $\vec{v}_{\mathrm{rel}}$ are the relative position and velocity vectors for particles A and B at the current as well as the next timestep:
\begin{eqnarray}
\vec{r}_{\mathrm{rel}} (t) = \vec{r}_A (t) - \vec{r}_B (t) , \: \: \: \vec{v}_{\mathrm{rel}} (t) = \vec{v}_A (t) - \vec{v}_B (t) 
\end{eqnarray}
A negative value of $\chi$ indicates that within $\Delta t$, particles A and B reach their distance of closes approach and their paths cross. As a next step we determine the collision time via \cite{Sagert2013}:
\begin{eqnarray}
t_{c \: 1,2} = \frac{1}{ |\vec{v}_{\mathrm{rel}} |^2} \left[ -  (\vec{v}_{\mathrm{rel}} \cdot \vec{r}_{\mathrm{rel}} )^2 \pm \sqrt{ ( \vec{v}_{\mathrm{rel}} \cdot \vec{r}_{\mathrm{rel}} )^2  - |\vec{v}_{\mathrm{rel}} |^2 \left( | \vec{r}_{\mathrm{rel}} |^2 - ( r_{\mathrm{eff}, A} + r_{\mathrm{eff}, B} )^2 \right) } \right] .
\label{collision_time2}
\end{eqnarray}
If the distance of closest approach between particles A and B is larger than the sum of their effective radii $r_{\mathrm{eff}}$ the collision time $t_c$ has imaginary values and the interaction does not take place. These first two steps typically leave several potential interaction partners for one particle of interest. The final collision partner is then determined from this sample as the particle with the shortest distance to the particle of interest. After collision partners have been assigned to each other, we perform the scattering in the center of mass frame of each collision pair. Previous approaches showed that a limitation of interactions to unambiguously determined scattering partners can lead to a significant underestimate of the collision rate \cite{Kortemeyer95}. In our work, we allow different particles to have the same collision partner and perform the corresponding interactions sequentially during one time-step $\Delta t$ \cite{Sagert2013}. We find that typically all test-particles interact with each other when matter is in the hydrodynamic regime. Once all collisions have been performed each particle's position is updated according to :
\begin{eqnarray}
\vec{x}( t + \Delta t) = \vec{x}(t) + \vec{v}_{\mathrm{old}} t_{\mathrm{min}} + \vec{v}_{\mathrm{new}} (\Delta t - t_{\mathrm{min}}), \: \: \mathrm{with} \: \: t_{\mathrm{min}} = - \left( \vec{r_{\mathrm{rel}}} \cdot \vec{v_{\mathrm{rel}}} \right) / \left| \vec{v_{\mathrm{rel}}} \right|^2 . 
\end{eqnarray}
Hereby, $t_{min}$ is the time when particles reach their distance of closest approach. This is an improvement of the approach which was used in \cite{Sagert2013}. Large particle densities and small mean-free-paths can lead to negative collision times $t_c$ (see eq.(\ref{collision_time2})). In \cite{Sagert2013} this case was treated as an instantaneous collision. However, by allowing particles to reach their point of closest approach during $\Delta t$, we achieve a better localization of shock fronts which is important in e.g. astrophysical simulations of core-collapse supernovae.
\section{Sedov shock test}
\label{sedov_test}
Shock wave studies are very well suited to demonstrate a code's ability to model hydrodynamic behavior. To reach the continuum regime, i.e. $K_n \ll 1$, we set the particle mean-free-path to be very small with respect to the width of a scattering grid cell. Since analytic shock wave solutions are generally obtained for an ideal gas equation-of-state, they require a simple description of particle interactions in from of elastic scattering. Hereby, to facilitate a simple treatment of particle interactions we assign particle degrees of freedom $f$ via the number of dimensions in the simulation. This affects the equation-of-state of the studies matter via its heat capacity ratio $\gamma = 1 + 1/f$ which becomes $\gamma = 2$ for a two dimensional simulation and $\gamma \sim 1.67$ in the case of three dimensions. It has already been demonstrated that the kinetic approach can reproduce the Sod and Noh shock problems \cite{Sagert2013}. The current work is focused on the Sedov blast wave test. While the Sedov test is a standard test for hydrodynamic codes it has not been discussed for kinetic schemes. The test describes an expanding spherical shock front and is therefore of special interest for codes which aim to simulate astrophysical problems, such as core-collapse supernovae. The initial setup of the Sedov test is a simulation space filled with matter at uniform density $n_{\mathrm{in}}$ and vanishing pressure. An explosion is set up via a point-like deposition of energy $E_{\mathrm{blast}}$ \cite{Taylor50,Sedov59}. The result is a spherical shock wave that moves outwards leaving behind matter with decreasing densities towards the center. While the bulk velocity of matter has a radial dependence $v_b \sim r/t$, with $t$ being the time and $r$ the radial distance from the center of the simulation space, the position of the blast wave can be written as:
\begin{eqnarray}
r_{\mathrm{shock}} = \left( E_{\mathrm{blast}} \: t^2 / \alpha \: n_{\mathrm{in}} \right)^{1/(2+d)} ,
\end{eqnarray}
with a peak density of $n(r) = n_{\mathrm{in}} \: (\gamma + 1)/(\gamma - 1)$. Here, $d$ is the number of dimensions in the simulation and $\alpha$ is a constant of the order one. 
\begin{figure}[h]
\begin{minipage}{14pc}
\includegraphics[width=14pc]{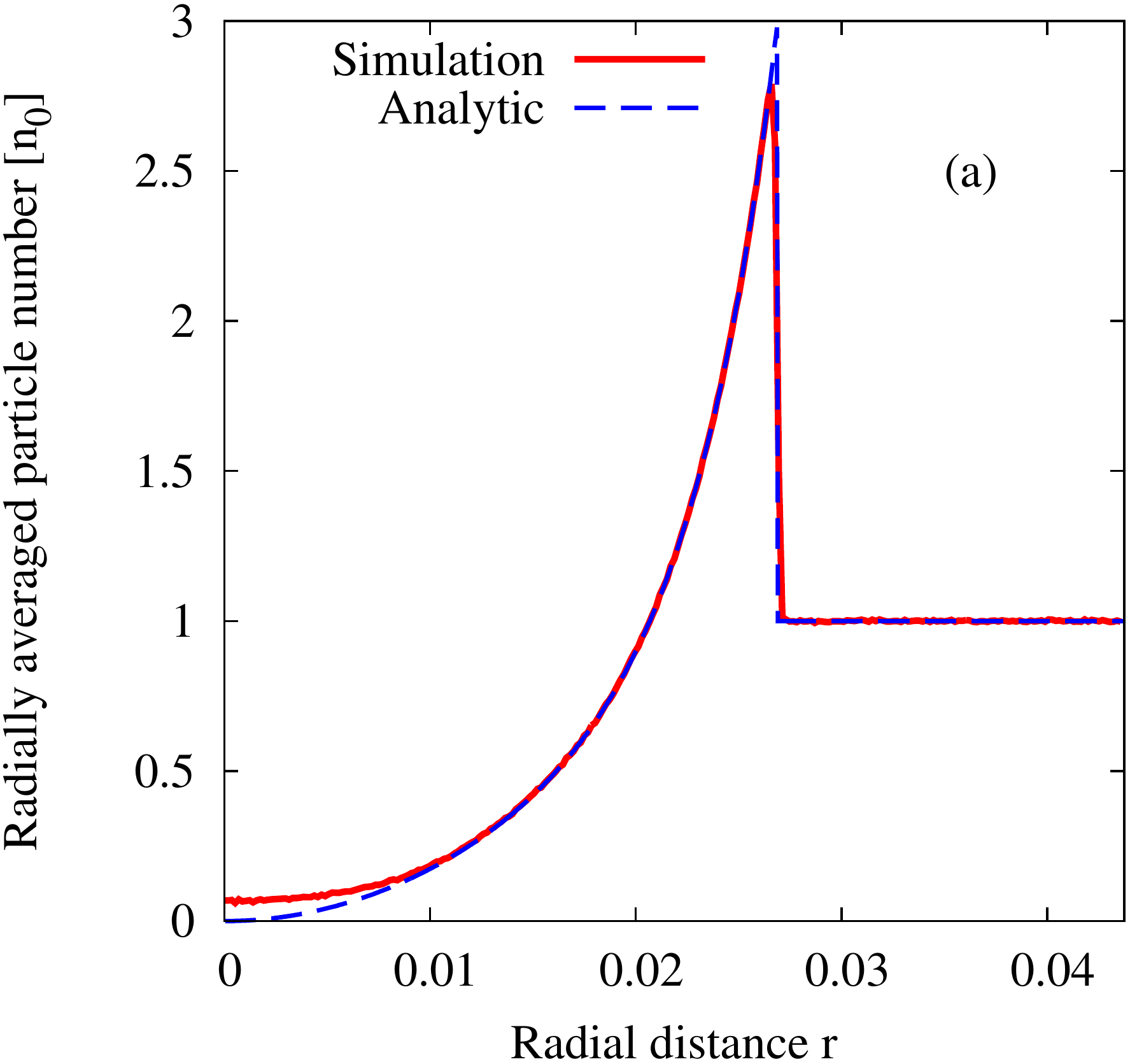}
\end{minipage}\hspace{4pc}%
\begin{minipage}{14pc}
\includegraphics[width=14pc]{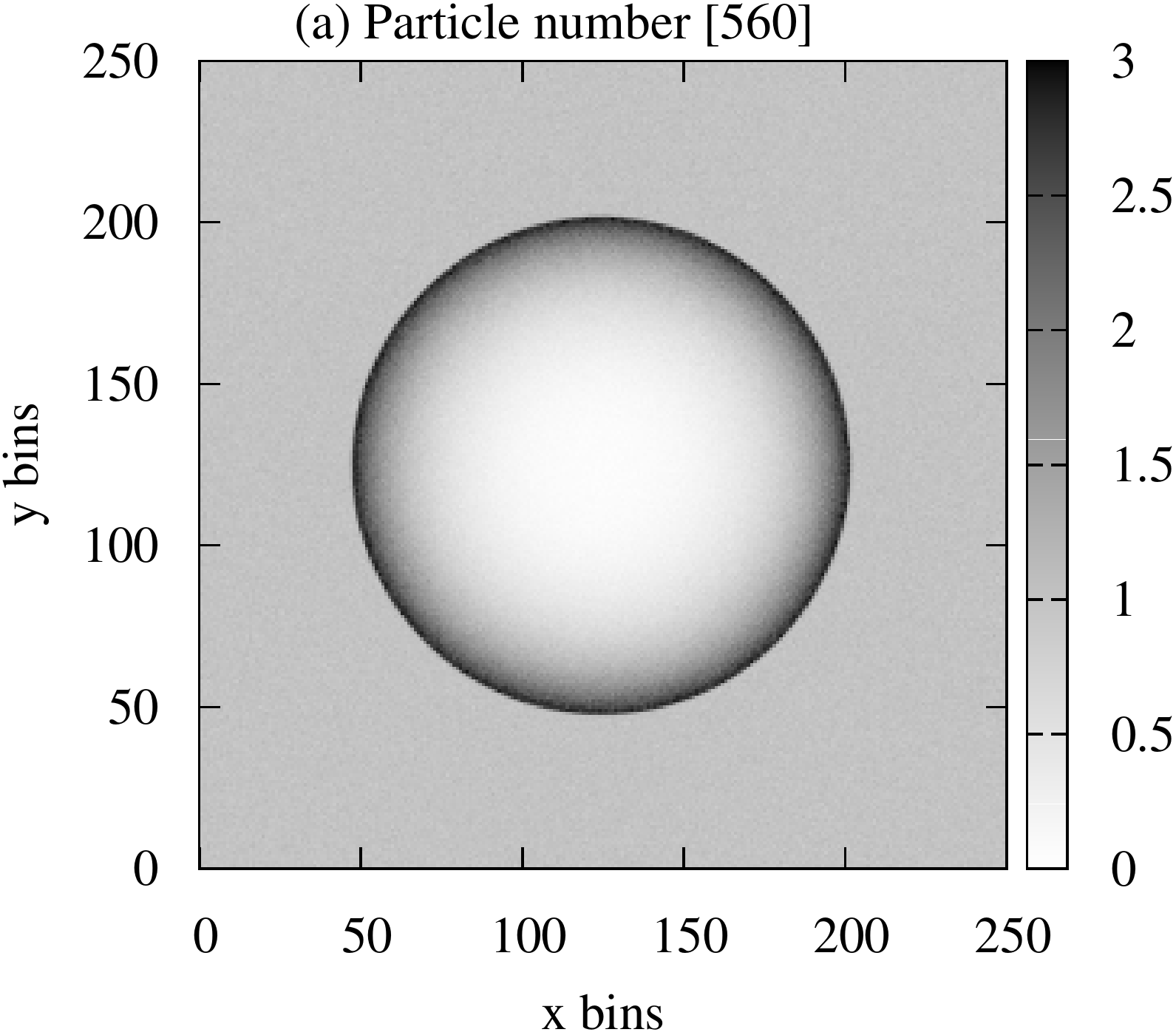}
\end{minipage}
\begin{minipage}{14pc}
\includegraphics[width=14pc]{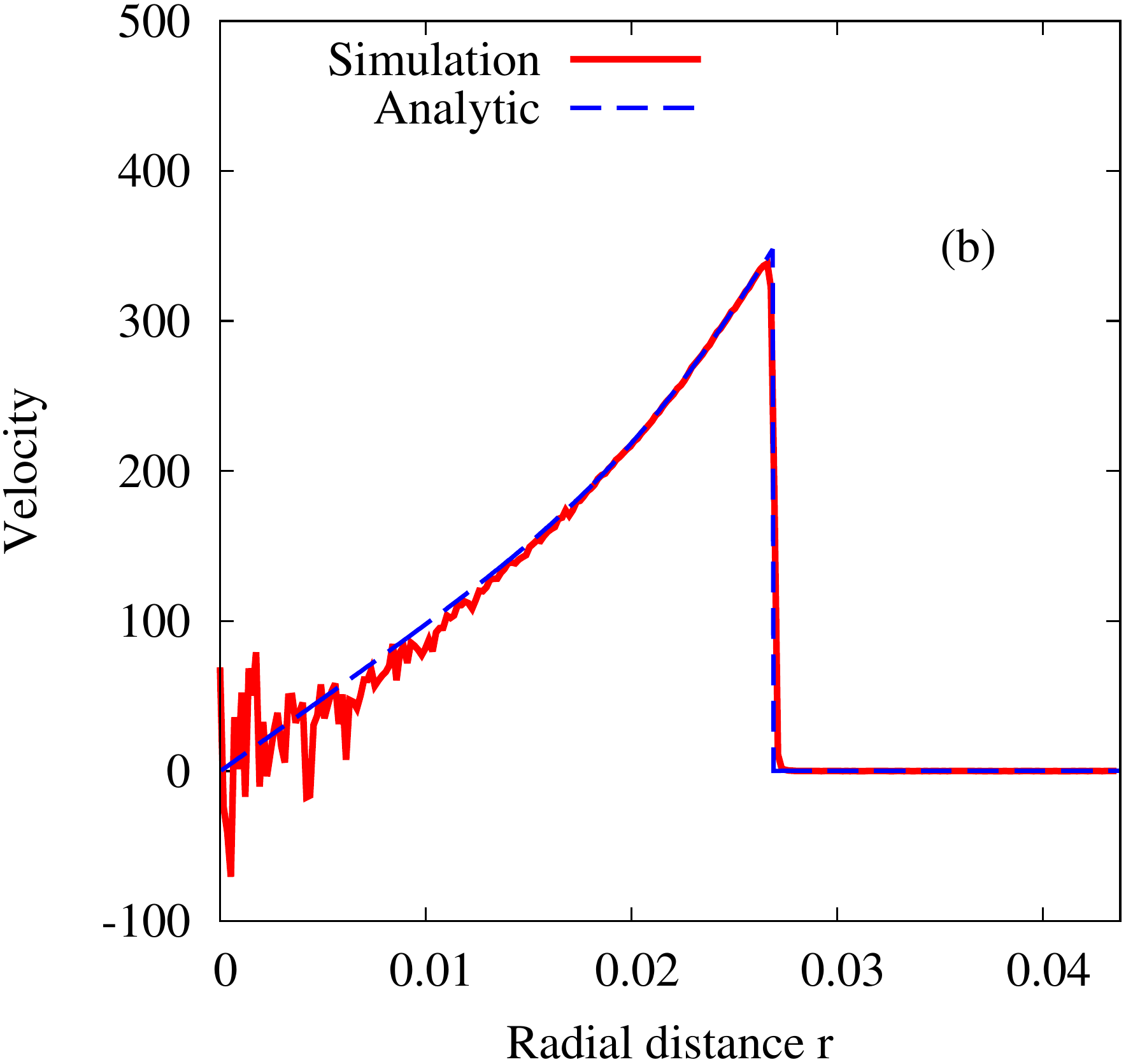}
\end{minipage}\hspace{4pc}%
\begin{minipage}{14pc}
\includegraphics[width=14pc]{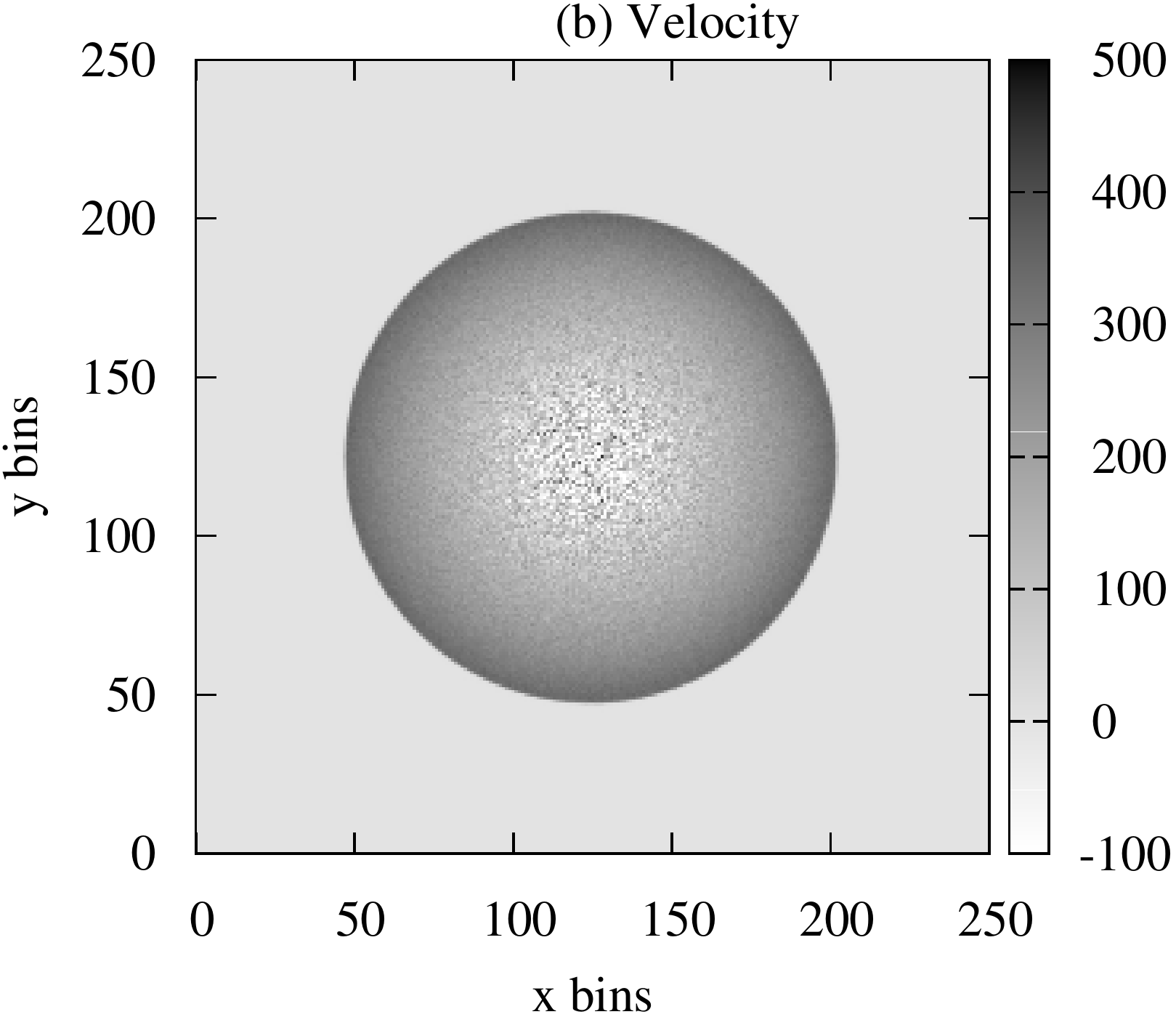}
\end{minipage} 
\begin{minipage}{14pc}
\includegraphics[width=14pc]{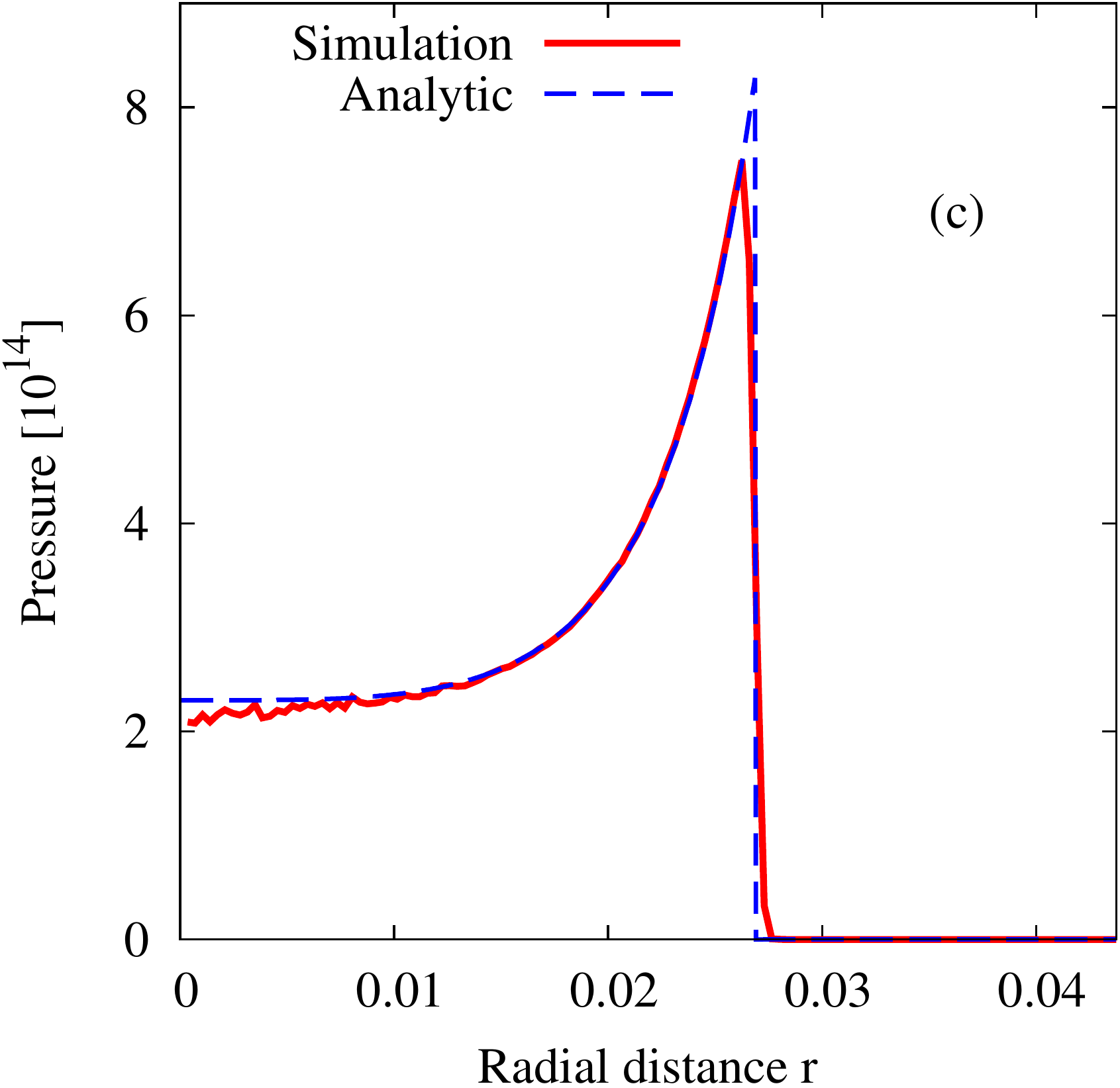}
\caption{\label{sedov1}Radial profiles of the particle density (a), velocity (b), and pressure (c) in the Sedov test}
\end{minipage}\hspace{4pc}%
\begin{minipage}{14pc}
\includegraphics[width=14pc]{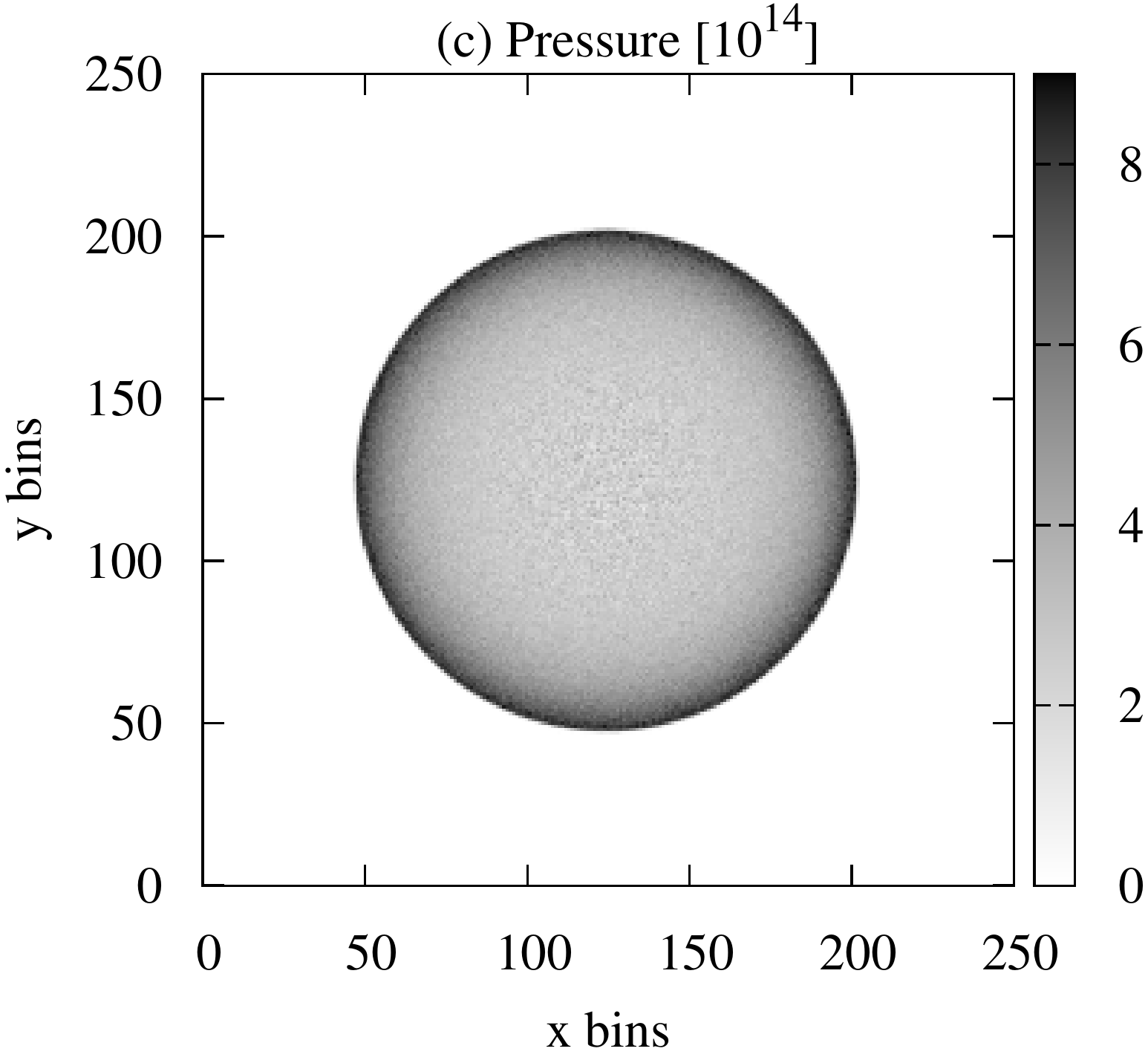}
\caption{\label{sedov2}Number of particles (a), velocity (b), and pressure (c) per bin in the Sedov test}
\end{minipage} 
\end{figure}
As the shock wave moves away from the center it leaves behind matter at vanishingly low densities. This can be a challenging problem for simulations operating with a finite number of particles since densities cannot become arbitrarily small \cite{Rosswog07}. Furthermore, as in all numerical studies, a general problem is the finite spatial resolution of a simulation which typically leads to a broadening of shock fronts. For the Sedov test, the latter impacts the density profile reducing its peak values. We perform the Sedov test in two dimensions with $N = 3.5\times 10^7$ test-particles distributed randomly in $250\times 250$ bins over $-0.04375 \leq x,y \leq 0.04375$. The blast wave is initialized around $x=y=0$ with a radius of $r_{\mathrm{blast}} = \Delta x$. The mean-free-path is chosen to $\lambda = 0.001 \: \Delta x$. While the particle velocities outside the blast region are initialized with absolute values of $v_{\mathrm{in}} = 3$, we deposit $E_{\mathrm{blast}}$ by setting $v_{\mathrm{in}} = 37180$ for particles with radial distance $r \leq r_{\mathrm{blast}}$. For this study we chose an adaptive timestep with $\Delta t = \Delta x / v_{max} $, whereas $v_{max}$ is the maximum absolute particle velocity at the current timestep. With a number of particles in the blast region of $1840$ the resulting explosion energy is $E_{\mathrm{blast}} \sim 1.27176 \times 10^{12}$. Figures \ref{sedov1}(a)-(c) show the the density, radial velocity, and pressure profiles at timestep 420 with $t = 2.57877 \times 10^{-5}$, averaged over $r$ together with the analytic solution. The latter is obtained by the publicly available \textit{sedov3.f} code \cite{Fryxell_riemann}. Figures \ref{sedov2}(a)-(c) show the particle numbers, bulk velocity, and pressure per bin in the simulation space at the same timestep. We find that the density of matter at the shock front is slightly reduced. However, as previously explained, this is an expected result of simulation due to the finite spatial resolution. The particle number in the center of the simulations has higher values than the analytic solution which can be attributed to the finite minimal number of test-particles in one cell. The small value of the latter on the other hand leads to large fluctuations in radial velocity at the center of the simulation as seen in Figs. \ref{sedov1}(b) and \ref{sedov2}(b) as well as in the pressure profile shown in Fig. \ref{sedov1}(c). The pressure is obtained from the stress tensor (see \cite{Sagert2013} for more details) via a radial average of the pressure per bin, shown in Fig. \ref{sedov2}(c). Overall, we see a very good agreement between the analytic solution and the kinetic simulations. 
\section{Summary}
In this work we present a transport code which combines the computational advantages of the Direct Simulation Monte Carlo technique with spatial resolution of the Point-of-Closest-Approach method to solve the collision integral of the Boltzmann equation. This enables us to simulate systems with a large number of test-particles in the hydrodynamic regime with a high spatial accuracy. To ensure that the hydrodynamic regime can be successfully reproduced we perform shock wave studies whereas the present work is focused on the Sedov shock wave test. To facilitate the comparison of our simulation to analytic solutions of shock wave propagation, we model particle interactions via elastic scatterings with the number of degrees of freedom given by the number of dimensions in the simulation. The current work verifies that our code is able to handle matter in the hydrodynamic regime and reproduce the Sedov shock with high accuracy. To our knowledge, there are no particle methods that addressed the Sedov shock test and our work represents the first of such studies. In the future we aim to apply our code to study the dynamics of core-collapse supernovae and matter in inertial confinement fusion capsules. 

\section*{References}
\providecommand{\newblock}{}

\end{document}